\documentclass[useAMS]{mn2e}
\usepackage{graphicx}
\usepackage[authoryear]{natbib}

\title[Read-out streaks in the XMM-OM]{The calibration of read-out-streak photometry in the XMM-Newton Optical Monitor and the construction of a bright-source catalogue}

\author[M.J. Page et al.]
{M.J. Page$^{1}$, 
N. Chan$^{1}$,
A.A. Breeveld$^{1}$, 
A. Talavera$^{2}$,
V. Yershov$^{1}$,
T. Kennedy$^{1}$,
\and N.P.M. Kuin$^{1}$,
B. Hancock$^{1}$, 
P.J. Smith$^{1}$, 
M. Carter$^{1}$\\ 
\\
$^{1}$Mullard Space Science Laboratory, University College London,
Holmbury St Mary, Dorking, Surrey, RH5 6NT, UK\\
$^{2}$XMM-Newton Science Operations Centre, ESA, Villafranca del Castillo,
Apartado 78, 28691, Villanueva de la Ca\~{n}ada, Spain\\
}

\begin{document}

\date{Accepted ----. Received ----; in original form ----}

\pagerange{\pageref{firstpage}--\pageref{lastpage}} 
\pubyear{2013}
\maketitle

\label{firstpage}

\begin{abstract}
  The dynamic range of the {\em XMM-Newton} Optical Monitor (XMM-OM)
  is limited at the bright end by coincidence loss, the superposition
  of multiple photons in the individual frames recorded from its
  micro-channel-plate (MCP) intensified charge-coupled device (CCD)
  detector. One way to overcome this limitation is to use photons that
  arrive during the frame transfer of the CCD, forming vertical
  read-out streaks for bright sources. We calibrate these read-out
  streaks for photometry of bright sources observed with XMM-OM. The
  bright source limit for read-out streak photometry is set by the
  recharge time of the MCPs. For XMM-OM we find that the MCP recharge
  time is $5.5\times 10^{-4}$~s. We determine that the effective
  bright limits for read-out streak photometry with XMM-OM are
  approximately 1.5 magnitudes brighter than the bright source limits
  for normal aperture photometry in full-frame images.  This
  translates into bright-source limits in Vega magnitudes of UVW2=7.1,
  UVM2=8.0, UVW1=9.4, U=10.5, B=11.5, V=10.2 and White=12.5 for data
  taken early in the mission. The limits brighten by up to 0.2
  magnitudes, depending on filter, over the course of the mission as
  the detector ages. The method is demonstrated by deriving UVW1 
  photometry for the symbiotic nova RR Telescopii, and the new 
  photometry is used to constrain the e-folding time of its decaying 
  UV emission. Using the read-out 
  streak method, we obtain
  photometry for 50 per cent of the missing UV source measurements in
  version 2.1 of the XMM-Newton Serendipitous UV Source Survey
  (XMM-SUSS~2.1) catalogue.
\end{abstract}


\section{Introduction}
\label{sec:introduction}

The {\em XMM-Newton} Optical Monitor \citep[XMM-OM;][]{mason01} is a
30~cm Ritchey Chr\'{e}tien telescope working at ultraviolet and
optical wavelengths. It is co-aligned with the X-ray mirrors on the
European Space Agency's {\em XMM-Newton} observatory. A similar
instrument, the Ultraviolet/Optical Telescope \citep[UVOT;][]{roming05} 
is carried on NASA's {\em Swift} gamma-ray-burst observatory.

The XMM-OM detector is a micro-channel plate (MCP) intensified charge
coupled device (CCD) \citep[MIC;][]{fordham89}. Incoming photons first
encounter a multi-alkali S20 photo-cathode, where they are converted
into one or two electrons which are then proximity focused onto a
series of MCPs. The MCPs multiply the number of incoming electrons by
a factor of $10^{6}$. The resulting electron cloud then encounters a
phosphor screen, which converts the signal back into photons. A
tapered block of optical fibres then transmits the photons to a
frame-transfer CCD with a science area of 256~$\times$~256 pixels. 
A single photon incident on
the photo-cathode results in a large splash of photons over several
pixels of the CCD. Photon splashes are detected in individual CCD
frames and centroided by the onboard electronics to a precision of one
eighth of a CCD pixel, equivalent to 0.5 arcsec on the sky.

When two or more incident photons give rise to overlapping photon
splashes on the CCD during a single CCD frame, they are recorded as a
single incident photon. This phenomenon is known as coincidence loss
\citep{fordham00b}. When the arrival rate of photons from an
astronomical source is small compared to the frame rate of the CCD,
coincidence loss is unimportant and the response of the instrument is
linear. Coincidence loss leads to an increasingly non-linear response
from the detector as the incident photon rate approaches the frame
rate of the CCD. For XMM-OM, the non-linearity is calibrated, and
photometry can be corrected in ground processing for incident photon
rates up to 3.6 times the CCD frame rate. 
Photometric uncertainty becomes large as this limit is approached 
\citep{kuin08}. Beyond this limit, the detected photon rate saturates 
at approximately the frame rate of the CCD and photometry is not 
recoverable via 
the usual aperture photometry method.

One way to overcome this limitation to the dynamic range of the
instrument is to use the photons that arrive during the frame transfer
of the CCD. At the end of each CCD frame, the CCD image is shifted to
the frame store area for read out. Photons that arrive during the
frame transfer (i.e. while the image is being moved downwards towards 
the frame store) are displaced in the vertical direction. Bright sources
therefore show vertical streaks of displaced photons which we will
refer to hereafter as read-out streaks. The frame transfer is
sufficiently fast that coincidence-loss is greatly reduced in the
read-out streak compared to the direct image.  \citet{page13} showed
that the read-out streaks in {\em Swift} UVOT images can be used for
photometric measurements, and demonstrated a photometric precision of
0.1 mag. They found that the bright limit for UVOT photometry could be
decreased by 2.4 magnitudes for full-frame images using this method,
with coincidence-loss within the MCPs rather than the CCD dictating
the bright limit for read-out-streak photometry.

In this paper we test and calibrate read-out-streak photometry for the
XMM-OM, before applying the method to construct a catalogue of
photometry for bright sources observed with XMM-OM.  In
Section~\ref{sec:instrument} we describe the basic principles of
photometry using read-out streaks. In Section~\ref{sec:method} we
describe the data and methods used to calibrate the recharge time of
the MCPs and verify read-out streak photometry against sources of
known brightness. The results are presented in
Section~\ref{sec:results}. In Section~\ref{sec:rrtel} we demonstrate 
the technique by deriving XMM-OM UVW1 photometry for the symbiotic nova RR 
Telescopii and examining its long term photometric evolution. 
In Section~\ref{sec:suppcat} we describe
the construction of a catalogue containing photometry of sources which
exceed the brightness limit of the XMM-SUSS~2.1 catalogue. Our
conclusions are presented in Section~\ref{sec:conclusions}.

Unless otherwise stated, magnitudes are given in the Vega system.

\section{Principles of photometry using read-out streaks}
\label{sec:instrument}

\begin{figure*}
\begin{center}
\includegraphics[width=140mm, angle=270]{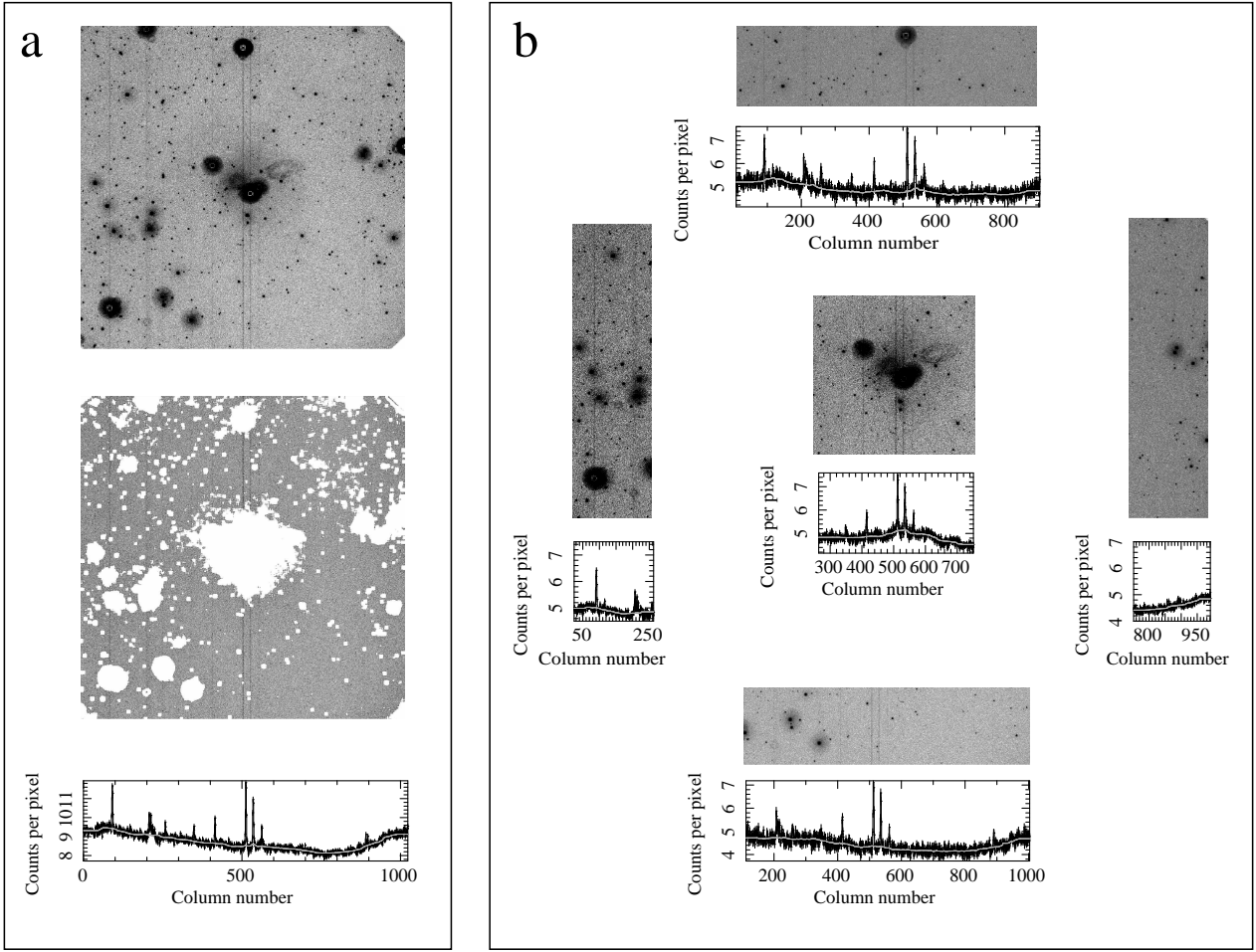}
\caption{Illustration of the read-out-streak and measurement process. Panel (a) shows a full-frame image before (top) and after (middle) masking of sources. In the graph at the bottom of panel (a), the black line shows the mean counts per pixel for each column of the masked image, while the grey line shows the background estimate (see Section \ref{sec:measurement}). The read-out streaks appear as positive spikes above the background estimate. Panel (b) shows the same field imaged in the default imaging mode (also known as `Rudi-5' mode) in which 5 separate images are taken to cover the field of view, together with the counts per column and background levels for each of the 5 images after masking.
}
\label{fig:streakdiagram}
\end{center}
\end{figure*}

The XMM-OM and {\em Swift} UVOT detectors were built to identical
specifications, so the principles of read-out streak photometry are
much the same in the two instruments and the description in
Section 2 of \citet{page13} is valid for 
XMM-OM as well. However, the two
instruments are operated rather differently. Whereas UVOT has only
three hardware-window modes and three corresponding frame times,
XMM-OM routinely uses a variety of hardware windows to image different
parts of the field of view (see Fig.~\ref{fig:streakdiagram} for an example), 
and so a variety of frame times. The
mathematical description that follows therefore differs from that
given in \citet{page13} in that it is valid for any frame time, rather
than for specific values. 

During frame transfer, charge is shifted 290 rows downwards from the
imaging portion of the CCD to the frame store area at a rate of 1 row
per $6\times 10^{-7}$~s, for a total transfer time of $1.74\times 10^{-4}$~s. 
Photons arriving
during this frame-transfer time form the read-out streak. Following
\citet{page13}, we base our read-out streak photometry on a
measurement in a $16 \times 16$ unbinned image pixel aperture, which
corresponds to $2 \times 2$ CCD pixels, and therefore an effective
exposure time corresponding to the transfer time of 2 CCD rows,
$1.2\times 10^{-6}$~s. For a frame time $t_{F}$, 
in which an image is accumulated
and then transferred, the ratio $S$ of the exposure time in the static
image to the read-out streak is therefore
\begin{equation}
S=\frac{t_{F} - 1.74 \times 10^{-4}}{1.2 \times 10^{-6}}
\label{eq:S}
\end{equation}

The zeropoints for read-out streak photometry are related to the normal 
imaging zeropoints by the ratio of equivalent exposure time $S$, such that
\begin{equation}
Z_{s} = Z_{i} -2.5 \log_{10} (S)
\label{eq:zeropoints}
\end{equation}
where $Z_{s}$ is the zeropoint for read-out-streak photometry and $Z_{i}$ is 
the imaging zeropoint.

A complication arises in that imaging zeropoints for XMM-OM, given in
the Current Calibration File
(CCF)\footnote{http://www.cosmos.esa.int/web/xmm-newton/current-calibration-files}
are given for 6 or 17.5 arcsec radius apertures, depending on filter
\citep{talavera11}, which have somewhat larger footprints than the
$16\times 16$ aperture used for the read-out streak
photometry. Therefore we have converted the published
zeropoints\footnote{calibration file OM$\_$COLOURTRANS$\_$0010.CCF} to
those which would be appropriate for a 5 arcsecond radius aperture
using the encircled energy fractions\footnote{calibration file
  OM$\_$PSF1DRB$\_$0010.CCF} for low-count-rate sources in the CCF to 
match better the
read-out-streak aperture, and to maintain consistency with the
approach in \citet{page13}. As the CCF does not contain a suitable
encircled energy fraction calibration for the White filter, we have
used the encircled energy fraction for the $U$ filter for this
purpose.  The adjusted imaging zeropoints, appropriate for use as
$Z_{i}$ in Equation~\ref{eq:zeropoints} are given in Table
\ref{tab:zeropoints}.

\begin{table}
\caption{XMM-OM imaging Vega and AB zeropoints scaled to a 5 arcsec radius aperture, suitable for use as $Z_{i}$ in Equation~\ref{eq:zeropoints}. For the White filter we do not provide an AB zeropoint because the CCF does not contain an AB zeropoint for this filter.}
\label{tab:zeropoints}
\begin{tabular}{lcc}
Filter&Vega zeropoint&AB zeropoint\\
           &(mag)&(mag)\\
\hline
&&\\
V& 17.9501  &17.9098\\
B& 19.2481  &19.0629\\
U& 18.2408  &19.1705\\
UVW1& 17.1247  &18.4871\\
UVM2& 15.7004  &17.3400\\
UVW2& 14.8200  &16.5252\\
White& 20.2370  &-\\
\hline
\end{tabular}
\end{table}

The dead time of the final-stage MCP is likely to limit the read-out
streak photometry at the bright end through a second stage of
coincidence loss that is independent of the CCD 
\citep{fordham00a, page13}. 
The expected count rate after coincidence loss is given by 
Equation 3 of \citet{page13} which is reproduced here:
\begin{equation}
R_{o} = \frac{1 - e^{-(S\ R_{i}\ \tau_{MCP})}}{S\ \tau_{MCP}} 
\label{eq:coiloss}
\end{equation}
where $R_{o}$ is the count rate observed from the read-out streak 
in a $16 \times 16$ pixel aperture, $R_{i}$ is incident count rate
that would be observed if there were no coincidence loss in the MCP,
and $\tau_{MCP}$ is the MCP recharge timescale.

Rearranging Equation~\ref{eq:coiloss}, we obtain the following 
equation for $R_{i}$:
\begin{equation}
R_{i} = -\frac{\log_{e}(1.0 - (S\ R_{o}\ \tau_{MCP}))}{S\ \tau_{MCP}}
\label{eq:coicorr}
\end{equation}
The gradual decline in XMM-OM sensitivity with time will affect read-out 
streak photometry in the same way as it affects normal aperture photometry. 
Therefore, $R_{i}$ should be corrected by the time-dependent sensitivity 
correction\footnote{calibration file OM$\_$PHOTTONAT$\_$0004.CCF} 
which is contained within the CCF. 
If $C$ is the time-dependent sensitivity correction factor and $R_{c}$ is 
the read-out-streak count rate after correction, then
\begin{equation}
R_{c}=C\times R_{i}
\label{eq:tdscorr}
\end{equation}
and $R_{c}$ can be used with the appropriate zeropoint $Z_{s}$ to 
obtain a magnitude $m$:
\begin{equation}
m=-2.5\log_{10}(R_{c}) + Z_{s}
\label{eq:mag}
\end{equation}


\section{Methods}
\label{sec:method}

\subsection{Input Catalogue and Dataset}
\label{sec:caldata}

In order to calibrate and qualify read-out-streak photometry for
XMM-OM, we must measure the read-out streaks for sources of known (or
predictable) brightness in one or more XMM-OM photometric
passbands. The most reliable and widely available photometry for stars
in the magnitude range relevant for read-out-streak photometry comes
from the Tycho-2 catalogue \citep{esa97}. 
Tycho-2 is a catalogue of stars for which
photometry was recorded by the star mapper on ESA's {\em Hipparcos}
satellite. Tycho-2 contains photometry in two bandpasses, $V_{T}$ and
$B_{T}$, which cover similar but not identical wavelength ranges as
the $V$ and $B$ bands of XMM-OM. 

In order to obtain a suitable dataset, we correlated the XMM-SUSS~2.1
catalogue \citep{page12}\footnote{updated information with respect to
  \citet{page12} relevant to version 2.1 of the XMM-SUSS is available
  from
  http://www.ucl.ac.uk/mssl/astro/space$\_$missions/xmm-newton/xmm-suss2}
with the Tycho-2 catalogue using a matching radius of 5 arcsec. In
order to avoid crowded fields we excluded matches within 20 degrees of
the Galactic plane. Matches were restricted to Tycho-2 sources with
$V_{T} \le 12$ mag which have the most reliable Tycho-2
photometry. Transformation from Tycho-2 photometry to photometric
systems similar to that of the XMM-OM is known to be less reliable for
stars of spectral type M than for stars of earlier spectral type
\citep{page13, esa97}. It is difficult to differentiate M stars from
earlier types using Tycho-2 $B_{T} - V_{T}$ colour, so we obtained
near-infrared photometry by cross correlating our sample with the Two
Micron All Sky Survey \citep[2MASS;][]{skrutskie06}. Stars with
$V_{T}-J>3.0$ were then excluded from our catalogue because they are
likely to be M stars. It simplifies the analysis to have a common
frame time for all observations, so the dataset was further restricted
to Tycho-2 stars observed with full-frame XMM-OM observations, which
have a frame time of 11.0388~ms. For stars which have been repeatedly
observed by {\em XMM-Newton}, for each filter we used data from only a
single {\em XMM-Newton} pointing. Hence each star is represented by a
single XMM-OM measurement in any filter, and the level of systematic
error derived from the analysis will be representative of the
systematics occurring in individual {\em XMM-Newton} pointings.

\subsection{Transformation from the Tycho-2 to XMM-OM photometric system}
\label{sec:transformation}

\begin{figure}
\includegraphics[width=84mm]{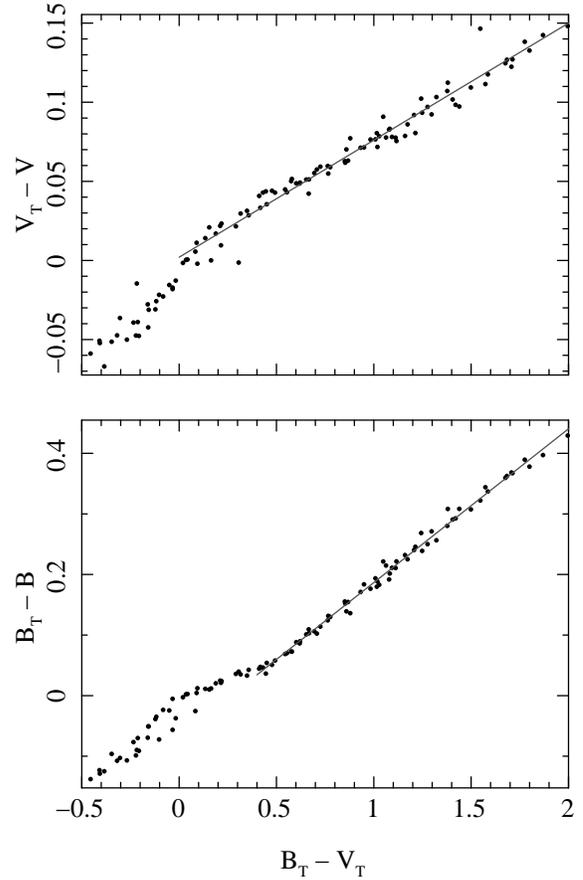}
\caption{Relations between XMM-OM $V$ and Tycho-2 $V_{T}$ magnitudes
  and between XMM-OM $B$ and Tycho-2 $B_{T}$ magnitudes with respect
  to Tycho-2 $B_{T}-V_{T}$ colour. The black dots correspond to
  synthetic photometry for stars in the \citet{pickles98} stellar
  library. The grey lines correspond to the linear relations adopted
  in this work for transformations between Tycho-2 and XMM-OM systems.
}
\label{fig:colours}
\end{figure}

To derive the transformation between Tycho-2 and XMM-OM photometric
systems, we generated synthetic photometry for the \citet{pickles98}
atlas of stellar spectra in both systems. Transformations involving
the Tycho-2 system are known to have a much larger scatter for M-type
stars than for earlier types \citep{esa97,page13}, and 
hence M star templates were not used
for the synthetic photometry. The colours relating $V$ and $B$ in the
XMM-OM system to $V_{T}$ and $B_{T}$ in the Tycho-2 system from the 
synthetic photometry are shown
in Fig.~\ref{fig:colours}.

The transformations between Tycho and XMM-OM are approximately linear
over a significant range in $B_{T}-V_{T}$. The following
transformations were obtained from least squared fits to the synthetic
photometry.

\begin{equation}
V = V_{T} - 0.002 - 0.074 (B_{T} - V_{T})  
\label{eq:vtransform}
\end{equation}
for $0 < (B_{T}-V_{T}) < 2$ and 
\begin{equation}
B = B_{T} + 0.0672 - 0.2538 (B_{T} - V_{T})  
\label{eq:btransform}
\end{equation}
for $0.4 < (B_{T}-V_{T}) < 2$. 
These relations are shown as solid lines in Fig.~\ref{fig:colours}. The rms 
scatter in $V_{T}-v$ and $B_{T}-b$ about these relations are 0.006 mag and 
0.009 mag respectively.

\subsection{Measurement of the read-out streaks}
\label{sec:measurement}

Read-out streaks were measured from raw XMM-OM images after correction
for modulo-8 noise by the standard {\em XMM-Newton} pipeline
processing. 

Prior to measuring the streaks, sources and bad pixels were masked from the
images. The first step was to mask bad pixels identified in the standard 
XMM-OM calibration 
file\footnote{calibration file OM$\_$BADPIX$\_$0005.CCF} within the CCF. 
Next, bright sources 
with count rates $>40$ counts~s$^{-1}$
were identified via a sliding-box search; bright sources are
surrounded by dark regions produced by coincidence loss, so a 24
arcsec radius region was masked around each bright source. Masking of
other sources was carried out on a column by column basis to prevent
the read-out streaks themselves being detected as sources. First, for
each column, the median brightness was calculated and any pixels
brighter than the median by more than $3~\sigma$, or by more than 3
counts if the median of the column is $\le 1$, were flagged. Next, the
column is smoothed with a 10 pixel box-car filter to aid the detection
of faint sources, and again any pixel which is brighter than the
median of the column by more than $3\sigma$, or 3 counts if the median
of the column is $\le 1$, were flagged. All pixels flagged in either
step were then masked.

Next, for each column the mean value for all non-masked pixels was
computed, and used to populate a single row of column brightness
values. For each pixel in this one-dimensional row, the median of the
pixel values within a 128 unbinned-pixel-wide box was used to define a
background, and read-out streaks were then detected using a sliding 16
unbinned-pixel cell. Examples showing the column brightness and
corresponding background for images taken in full-frame mode and the
default imaging mode are shown in Fig.~\ref{fig:streakdiagram}. For
streaks detected at $> 6~\sigma$, the count rate was summed over the
16-pixel interval and background subtracted. Next, the count rates
were multiplied by a factor of 16 to scale them from the equivalent of 
a single-pixel slice 
of the aperture to the full $16 \times 16$ pixel
aperture. Finally, the count rates of the read-out streaks were
corrected for the time-dependent sensitivity degradation of the XMM-OM
\citep{talavera11}.

The procedure just described differs from that described in Section
3.2 of \citet{page13} for UVOT read-out streaks only in two
aspects. First, no bad-pixel mask was employed in \citet{page13}, but
in the present work we have found that masking the bad pixels improves
the read-out streak detection and photometry towards the edges of the
images.  Second, whereas corrections for large scale sensitivity
variations of UVOT were applied in \citet{page13}, there are no such
corrections for XMM-OM.

Where multiple full-frame exposures have been taken through the same filter in
the same {\em XMM-Newton} observation, the read-out streak count rates
were averaged. To ensure good quality photometry for the calibration,
only measurements with a signal to noise ratio $>20$ were used. There are 
suitable measurements for 54 stars in the $V$ band and 25 stars in the $B$ band.

\section{Results}
\label{sec:results}

\begin{figure}
\includegraphics[width=84mm]{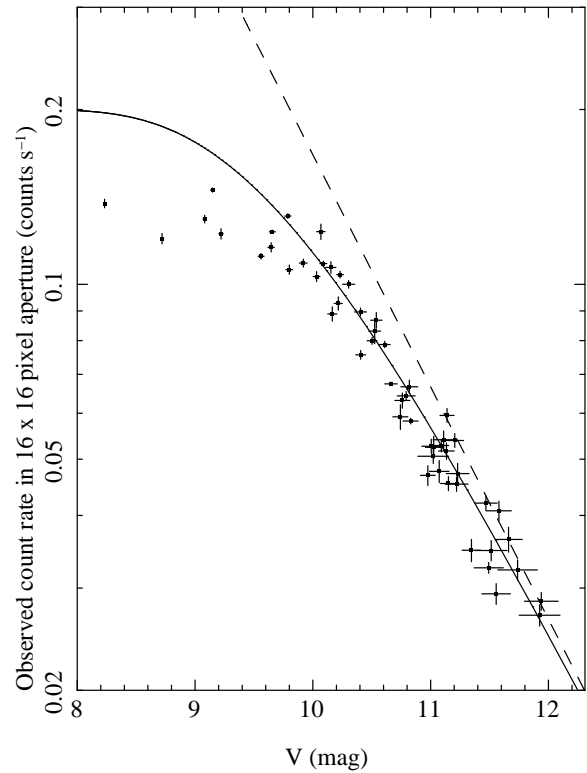}
\caption{Datapoints show the observed read-out-streak count rates in
  $16\times16$-pixel apertures (equivalent to $R_{o}$ in 
Equations \ref{eq:coiloss} and \ref{eq:coicorr}) of Tycho-2 stars 
against XMM-OM~$V$
  magnitude, derived from Tycho-2 photometry
  (equation~\ref{eq:vtransform}). The dashed line shows the expected
  relationship if there were no coincidence-loss in the MCPs, (i.e. if
  $R_{i}$=${R_o}$). The solid line shows the expected relationship for
  an MCP recharge time $\tau_{MCP}=5.5\times 10^{-4}$~s, as derived in
  Section~\ref{sec:results}.}
\label{fig:v_mag_rate}
\end{figure}

Figure \ref{fig:v_mag_rate} shows the count rates measured in the
read-out streaks against the predicted $V$ magnitudes for the Tycho-2
stars. The dashed line shows the predicted relationship if there were
no coincidence loss within the MCPs. While the measurements approach
the dashed line at faint magnitudes, the difference is large at
bright magnitudes, implying that, as expected, coincidence loss in the
XMM-OM MCPs has a significant effect on the read-out streaks of bright
sources. For sources brighter than $v=10$ mag the read-out streak
count rate shows only a weak dependence on source magnitude. 
To determine
the MCP recharge timescale, $\tau_{MCP}$, we performed a $\chi^{2}$
fit in which the observed count rate is related to the incident count
rate according to Equation~\ref{eq:coiloss} to the sources fainter
than $V=10$~mag in Fig.~\ref{fig:v_mag_rate}. We obtain a
best-fitting $\tau_{MCP}=5.5\pm0.1 \times 10^{-4}$~s.

\begin{figure}
\includegraphics[width=84mm]{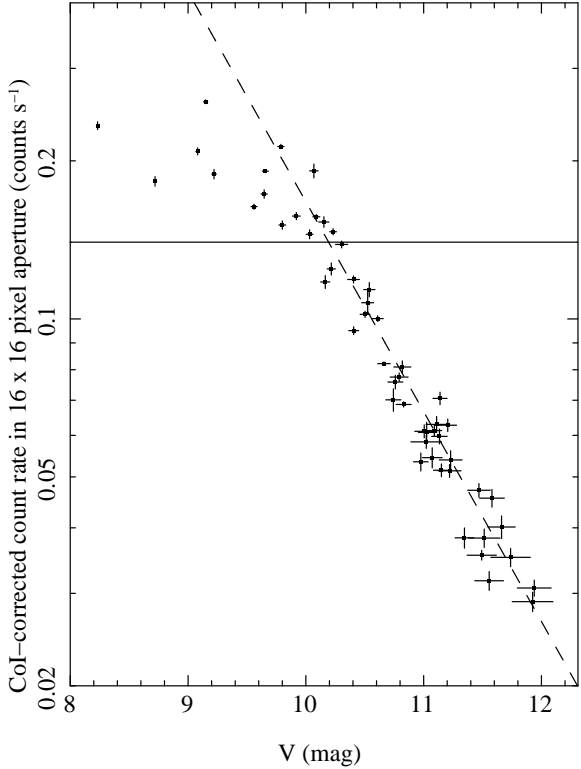}
\caption{Datapoints show the read-out-streak count rates in
  $16\times16$-pixel apertures of Tycho-2 stars in full-frame XMM-OM V-band images after correction for coincidence loss in the MCPs (equivalent to $R_{i}$ in 
Equations \ref{eq:coiloss} and \ref{eq:coicorr})  
against XMM-OM~$V$
  magnitude, derived from Tycho-2 photometry
  (Equation~\ref{eq:vtransform}). The dashed line shows the expected
  relationship. The solid line shows a count rate of 0.14 counts~s$^{-1}$, below which saturation effects are not important.}
\label{fig:v_mag_rate_coi}
\end{figure}

\begin{figure}
\includegraphics[width=84mm]{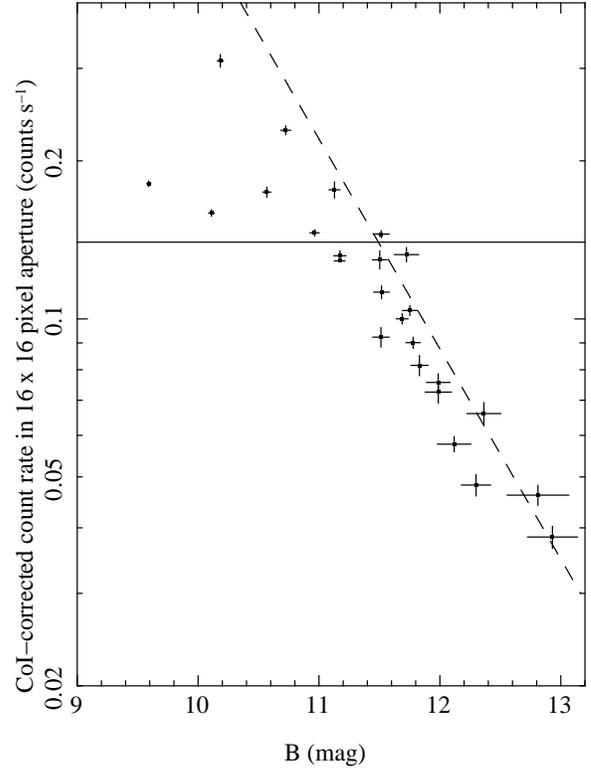}
\caption{Datapoints show the read-out-streak count rates in
  $16\times16$-pixel apertures of Tycho-2 stars in full-frame XMM-OM B-band images after correction for coincidence loss in the MCPs (equivalent to $R_{i}$ in 
Equations \ref{eq:coiloss} and \ref{eq:coicorr})  
against XMM-OM~$B$
  magnitude, derived from Tycho-2 photometry
  (Equation~\ref{eq:btransform}). The dashed line shows the expected
  relationship. The solid line shows a count rate of 0.14 counts~s$^{-1}$, below which saturation effects are not important.}
\label{fig:b_mag_rate_coi}
\end{figure}

The solid line in Fig.~\ref{fig:v_mag_rate} shows the expected
relation between V magnitude and count rate for the best fitting
value of $\tau_{MCP}$. The model reproduces the data well up to the
$V=10$~mag bright limit of the fit, but for $V<10$~mag the model and
data diverge systematically towards brighter magnitudes, with the
count rates falling below the expectations of the model for all
sources with $V<9.8$~mag. 
Furthermore, at the brightest magnitudes,
the statistical uncertainties in the measurements are small, and 
therefore we would
expect the scatter in the measured count rates to diminish towards
bright magnitudes as the count rates converge towards the saturation
limit. Instead, the small number of measurements at $V<9.5$~mag
exhibit significant scatter.
This scatter, combined with the relatively flat relation between 
count rate and magnitude preclude the use of read-out streaks for 
photometric measurement at the brightest magnitudes.
Similar deviations from the model
predictions at bright fluxes were observed by \citet{page13} in the
{\em Swift} UVOT, who suggested that charge bleeding in the CCD, or
positional dependencies in electron mobility in the MCPs, might be
responsible. It is interesting to note that $\tau_{MCP}=5.5\times
10^{-4}$~s is significantly longer than that found by \citet{page13}
for {\em Swift} UVOT, such that coincidence loss in the MCPs becomes
significant at a lower count rate in XMM-OM than in UVOT. The
bright limit beyond which the data deviate from the coincidence-loss
model is also shifted to fainter magnitudes in XMM-OM compared to
UVOT. This finding suggests that it is the behaviour of the MCPs, rather
than charge bleeding in the CCD, which is responsible for the break
down of the model at bright magnitudes.

Correcting the observed read-out-streak count rates for coincidence
loss in the MCPs using Equation~\ref{eq:coicorr} and
$\tau_{MCP}=5.5\times 10^{-4}$~s, we obtain
Fig.~\ref{fig:v_mag_rate_coi}. A close correlation is seen between
count rate and magnitude for coincidence-loss corrected count rates
$R_{i} < 0.14$~counts~s$^{-1}$. Fig.~\ref{fig:b_mag_rate_coi} shows
the coincidence-loss-corrected count rates for the $B$ band compared
to $B$ magnitudes derived from Tycho-2 photometry. There are fewer $B$
measurements, and the Tycho-2-derived $B$ magnitudes have larger
uncertainties than those for the $V$ band. In addition, read-out
streaks typically have larger statistical uncertainties in $B$ than in
$V$ because XMM-OM $B$ images typically have higher background levels
than $V$ images. Consistent with the findings from $V$-band
photometry, saturation effects associated with the MCPs are only seen
for $R_{i} > 0.14$~counts~s$^{-1}$.

\subsection{Photometric scatter}
\label{sec:scatter}

\begin{figure}
\includegraphics[width=84mm]{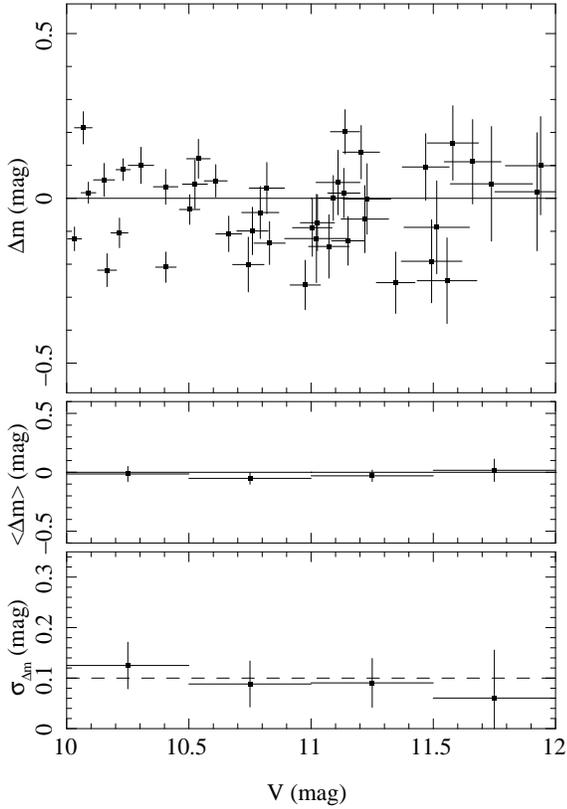}
\caption{Top panel: differences $\Delta m$ between the $V$ magnitudes
  obtained from Tycho-2 and the $V$ magnitudes obtained from the
  coincidence-loss corrected count-rates of the XMM-OM read-out
  streaks. Uncertainties are the quadrature sums of the errors on the
  count-rates and the Tycho-2 magnitudes. Middle and bottom panels:
  mean and dispersion of $\Delta m$ respectively in 0.5 magnitude
  bins. The dashed line in the bottom panel shows the best-fit
  dispersion, $\sigma_{\Delta~m}=0.10$, over the full $10<V<12$ magnitude 
  interval shown.}
\label{fig:vmag_delta_m}
\end{figure}

\begin{figure}
\includegraphics[width=84mm]{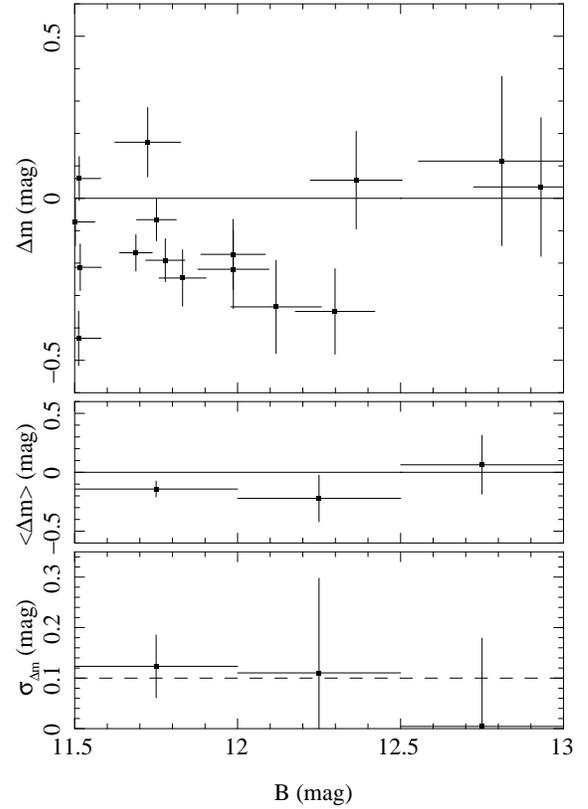}
\caption{Top panel: differences $\Delta m$ between the $B$ magnitudes
  obtained from Tycho-2 and the $B$ magnitudes obtained from the
  coincidence-loss corrected count-rates of the XMM-OM read-out
  streaks. Uncertainties are the quadrature sums of the errors on the
  count-rates and the Tycho-2 magnitudes. Middle and bottom panels:
  mean and dispersion of $\Delta m$ respectively in 0.5 magnitude
  bins. The dashed line in the bottom panel corresponds to a dispersion of 
  0.1 magnitudes. }
\label{fig:bmag_delta_m}
\end{figure}

We now examine the scatter of the read-out-streak photometry
measurements with respect to the Tycho-2 photometry to estimate the
level of photometric accuracy which can be achieved for XMM-OM using
read-out streaks. For UVOT, \citet{page13} found that systematic
errors limit the photometric accuracy of read-out-streak photometry to
0.1~mag. 

The top panels of Figs~\ref{fig:vmag_delta_m} and
\ref{fig:bmag_delta_m} show the differences between the
read-out-streak and Tycho-2 derived magnitudes ($\Delta~m$) for the
individual stars in $V$ and $B$. The uncertainties on $\Delta~m$ have
been computed by adding in quadrature the statistical uncertainty on
the read-out-streak photometry with the magnitude uncertainty given in
the Tycho-2 catalogue. The $V$ data in Fig.~\ref{fig:vmag_delta_m} are
of higher quality than the $B$ data in Fig.\ref{fig:bmag_delta_m},
both in terms of the precision of the individual measurements, and in
the number of measurements. We have used the method of
\citet{maccacaro88} to derive from these measurements in 0.5 magnitude
bins, maximum-likelihood estimates of the mean and intrinsic standard
deviation of the distribution of $\Delta~m$, which we have assumed to
be Gaussian. The middle panels of Figs~\ref{fig:vmag_delta_m} and
\ref{fig:bmag_delta_m} show the means, $\langle\Delta~m\rangle$, and
the bottom panels show the intrinsic standard deviations
$\sigma_{\Delta~m}$. The values of $\langle\Delta~m\rangle$ are all
consistent with 0, and no systematic trends are evident with
magnitude. The individual values of $\sigma_{\Delta~m}$ are less well
determined. For the $V$ band, applying the \citet{maccacaro88} method
over the full range of magnitudes we obtain $\langle\Delta~m\rangle
=-0.025\pm0.025$ and $\sigma_{\Delta~m}=0.10\pm0.02$. The individual
values of $\sigma_{\Delta~m}$ obtained in 0.5 magnitude bins in both
$V$ and $B$ bands are all consistent with $\sigma_{\Delta~m}=0.10$,
which represents the level of systematic uncertainty which must be
added in quadrature to the statistical uncertainties to reproduce the
distribution of $\Delta~m$. Thus photometry obtained from read-out
streaks should be considered to have a systematic uncertainty of
0.1~mag, in addition to the statistical uncertainty. It is interesting
to note that this level of systematic error, 0.1~mag, is identical to
that derived for UVOT read-out-streak photometry by \citet{page13}.


\subsection{Brightness limits for images which are not in full-frame mode}

From Figs~\ref{fig:v_mag_rate_coi} and \ref{fig:b_mag_rate_coi} we
identified a bright limit of $R_{i}=0.14$~counts~s$^{-1}$ for read-out
streaks in full-frame XMM-OM imaging, beyond which saturation of the
MCPs degrades the photometry. Different window configurations
correspond to different frame times, and so different exposure-time
ratios, $S$, according to Equation~\ref{eq:S}. For full-frame imaging
the frame time is 11.0388~ms, corresponding to $S=9054$.  The maximum
acceptable countrate will scale inversely with $S$. so for an
arbitrary frame time the bright-limit count rate for read-out streak
photometry is
\begin{equation}
R_{max} = \frac{1268}{S}
\label{eq:brightlimit}
\end{equation}

\section{A demonstration of read-out streak photometry: the near ultraviolet evolution of RR Telescopii}

\label{sec:rrtel}

The Nova RR Telescopii is a symbiotic binary which underwent a nova
outburst in 1944. In optical radiation it has been slowly dimming ever
since \citep{kotnik06}. Between 1978 and 1995 its ultraviolet flux
declined by a factor of 2--3, depending on wavelength
\citep{nussbaumer97}. Shortward of 1500~\AA, the UV flux of RR Tel is
thought to be dominated by the central star, while longward of
1800~\AA\ the UV flux is likely to be dominated by nebular emission
\citep{nussbaumer97}. In a study of the X-ray and ultraviolet emission
from RR Tel, based on an {\em XMM-Newton} observation taken in 2009,
\citet{gonzalez13} show that the 2009 XMM-OM photometry through the
UVW2 and UVM2 filters are consistent with an extrapolation of the
exponential decay seen in International Ultraviolet Explorer (IUE) 
data between 1978 and 1995. However,
\citet{gonzalez13} describe the XMM-OM UVW1 photometry as uncertain,
because it is subject to a large degree of coincidence loss. Indeed, in the
XMM-SUSS~2.1 catalogue there is no UVW1 magnitude for this source because 
it is flagged
as being too bright for a reliable photometric measurement in UVW1. 
This is unfortunate,
because it can be seen in \citet{gonzalez13} that the IUE-based
photometry corresponding to the UVW1 filter is of a higher
statistical quality than the photometry corresponding to UVW2 or
UVM2. Using the read-out streak technique, we are now able to
derive valid UVW1 photometry from the 2009 {\em XMM-Newton}
observation of RR Tel, and so can examine the UV photometric
evolution over a 31-year time base.

Photometry was derived from the 2009 XMM-OM UVW1 observation using the
read-out-streak technique described here. The read-out streak is present in
three of the five sub-exposures used to image the field around RR~Tel,
and is detected in all three sub-exposure with signal to noise ratios
of between 46 and 48. Thus the uncertainty on the photometry derived
from the read-out streak is dominated by the 0.1 mag systematic
term described in Section~\ref{sec:scatter}.

\begin{figure}
\includegraphics[width=60mm, angle=270]{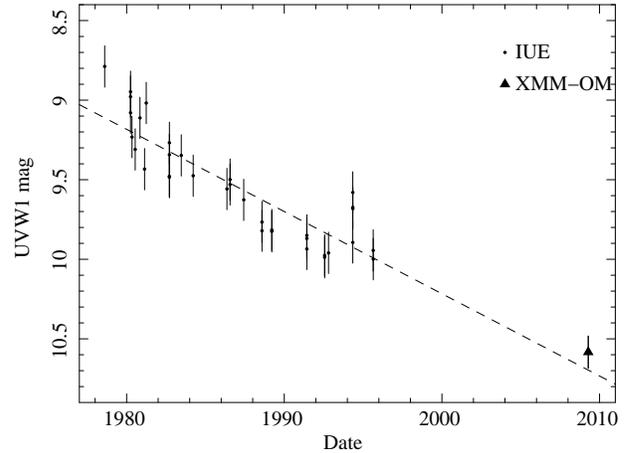}
\caption{Lightcurve of RR Tel in the UVW1 bandpass, using read-out streak photometry from XMM-OM and synthesized UVW1 photometry from IUE. The dashed line shows the best-fit exponential dimming model, which corresponds to an e-folding time of 19.4 years.}
\label{fig:rrtel}
\end{figure}

IUE large aperture, low-resolution spectra, reduced using the 
New Spectral Image Processing System \citep[NEWSIPS;][]{nichols96}, 
were retrieved from the Mikulski Archive for Space Telescopes.
UVW1 photometry was synthesized from the IUE 
spectra by integrating the product of IUE flux and the
the XMM-OM UVW1 response curve.  
The UVW1 passband extends further to the red than the spectral
coverage of IUE, which ends at 3350~\AA. In order to correct the IUE
synthesized photometry for the missing red flux, we made use of a
Hubble Space Telescope (HST) Space Telescope Imaging Spectrometer
(STIS) observation of RR~Tel from 2000, which provides full spectral
coverage throughout the UVW1 passband. From the STIS data, we
determine that the photometry of RR~Tel synthesized only to 3350~\AA\ 
is 0.12 mag fainter than the photometry synthesized over the full
bandpass.

The statistical uncertainties on the UVW1 photometry synthesized from
IUE spectra were initially propagated from the statistical
uncertainties contained within the NEWSIPS spectral files. Examination
of the scatter of the photometric data suggests that these
uncertainties are too small. In particular, the rms of the UVW1
photometry derived from four IUE observations taken within a single
day, 7 May 1994, is observed to be 0.13 mag, whereas the statistical
uncertainties derived from the spectra are only 0.01 mag. Therefore we
have adopted 0.13 mag as the statistical uncertainty on the
IUE-derived photometry.

Fig.~\ref{fig:rrtel} shows the XMM-OM UVW1 photometry together with
the corrected, synthesized UVW1 lightcurve from IUE. We have not
included the photometry derived from the HST STIS observation because
this was obtained from a much smaller aperture (0.2 arcsec) than the
IUE or XMM-OM photometry, and the spatial extent of the nebula is not
known. The dashed line shows the best-fit exponential decay model to
the IUE and XMM-OM UVW1 photometry. 
From the fit we derive an
e-folding time of $19.4\pm 1.2$ years, which is somewhat shorter than
the UVW1 e-folding time of $24\pm 1$~years 
derived by \citet{gonzalez13}, but consistent
with the e-folding times they derived for the shorter-wavelength UVW2
and UVM2 bands, which were $23\pm 5$~years and $21\pm 2$~years respectively.

\section{A supplementary catalogue of photometry for bright XMM-OM sources}
\label{sec:suppcat}

The
XMM-SUSS\footnote{http://www.cosmos.esa.int/web/xmm-newton/om-catalogue}
is a catalogue of sources detected in XMM-OM images and consists of
astrometric, photometric, and morphological information together with
quality flags to appraise the user of the validity of the measurements
\citep{page12}. The latest release of the catalogue, XMM-SUSS~2.1\footnote{http://www.ucl.ac.uk/mssl/astro/space$\_$missions/xmm-newton/xmm-suss2},
contains more than 4.3 million unique optical and UV sources. Sources
which exceed a threshold of 0.97 counts per frame in a given band are
deemed too bright for reliable photometry in that band. In such cases,
magnitudes are not provided in the catalogue and quality flags
indicate the bands for which the source is too bright. The
read-out-streak photometry method which has been described in this
paper permits photometry of sources up to 1.5 magnitudes brighter than
this threshold, and therefore offers the possibility to recover some of
the missing photometry for these bright sources. We have therefore
used read-out-streak photometry to produce a catalogue of photometric
measurements for bright sources to form a supplement to the main
XMM-SUSS~2.1 catalogue. In the following subsections we describe how we
constructed the supplementary catalogue and describe its basic
properties.

\subsection{Construction of the catalogue}
A pipeline was constructed to measure read-out streaks in XMM-OM
images and match them to stars in the XMM-SUSS which were too bright
for their photometry to be measured from the direct image. Such
sources can be identified in the XMM-SUSS~2.1 catalogue through having
quality flag number 11 (with numerical value of 2048) set in the
QUALITY$\_$FLAG column\footnote{see
  http://www.ucl.ac.uk/mssl/astro/space$\_$missions/xmm-newton/xmm-suss2
  for details of the XMM-SUSS~2.1 quality flags} corresponding to the
affected filter(s).  As for the XMM-SUSS, each individual {\em
  XMM-Newton} pointing, corresponding to a unique observation
identification number (OBSID) is treated independently in the
supplementary catalogue.  For each {\em XMM-Newton} observation
containing one or more saturated stars, we retrieved the standard
XMM-OM sourcelist and modulo-8-corrected raw-image products from the ESA
{\em XMM-Newton} Science Archive (XSA). Read-out streaks were measured
from the images using the software developed by \citet{page13}, with
bad pixels masked using the standard XMM-OM calibration
file\footnote{calibration file OM$\_$BADPIX$\_$0005.CCF} 
from the CCF. The columns
containing read-out streaks were then matched to the columns
containing saturated stars in the sourcelists. For XMM-OM observations
carried out in the default imaging mode (also known as `Rudi-5' mode)
in which 5 separate images are taken in a mosaic pattern to form a
full-field image, the read-out streaks often cross multiple images,
but the saturated star is typically found in only one. Therefore the
pipeline merges the individual sourcelists prior to matching the
read-out streaks to stars.

It is possible for more than one bright source to land on the same
column within the XMM-OM image, particularly in crowded fields near to
the Galactic plane, and in such cases multiple stars will contribute
to the read-out streak, invalidating the photometry. We have discarded
read-out-streak photometry for saturated stars in which one or more
additional stars with count rates of $>20$ counts~s$^{-1}$ are found
in the source list within eight raw columns of the saturated star. The
threshold of 20~counts~s$^{-1}$ corresponds to a maximum contaminating
contribution of $8$~per~cent 
to the read-out-streak of a saturated star.

We also identified cases in which two read-out streaks are detected in
close proximity (within 24 raw columns), and discarded any
corresponding read-out-streak photometry. Although this condition
occasionally identifies cases of bright sources lying on close-enough
columns that their read-out streak photometry is affected, its main
purpose in the pipeline is to identify stars which are far too bright
for read-out streak photometry but can not be identified as such from
the count-rates of their read-out streaks \citep[see Section 5.1 of
][]{page13}. Visual screening was used to reject other examples of
such stars which are too bright for read-out-streak photometry, but
which were not identified by their count rates or by the apparent
proximity of multiple read-out streaks.

During the construction of the catalogue it became evident that in
extremely crowded star fields, background determination becomes very
challenging. In extreme cases, the column-by-column background as
determined in the pipeline by the algorithm described in
Section~\ref{sec:measurement} shows large systematic deviations that
compromise the read-out streak measurement. From visual examination of
the background curves for fields containing different densities of
sources we have chosen a threshold of 2000 sources per full-frame
XMM-OM image; where the field is imaged through a sequence of windowed
exposures as in the default imaging mode, the sources are summed over all 
the exposures. We have
discarded read-out-streak photometry derived from images which exceed
this threshold.

\subsection{Properties of the catalogue}
\label{sec:properties}

\begin{figure*}
\includegraphics[width=150mm]{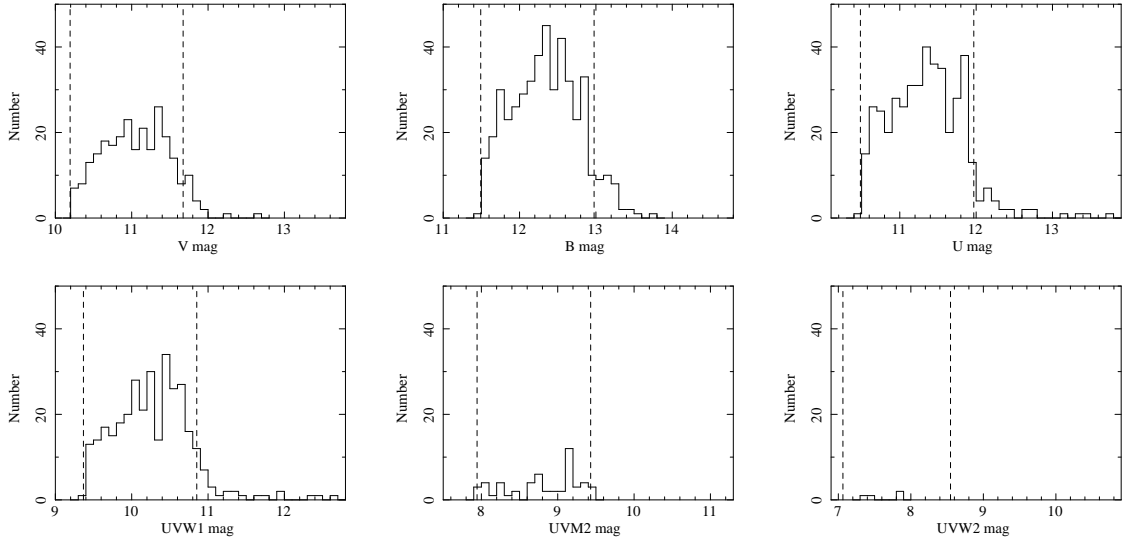}
\caption{Magnitude distributions of the supplementary catalogue
  entries for the six passbands used in the XMM-SUSS. The two dashed
  lines in each panel indicate the bright limits for read-out-streak
  photometry and aperture photometry.}
\label{fig:supp_mags}
\end{figure*}

Basic statistics for the supplementary catalogue are presented in
Table~\ref{tab:supplementary}. Overall, valid photometry is recovered
for 50 per cent of the UV measurements and 30 per cent of the optical
measurements, which are saturated in XMM-SUSS~2.1, using read-out
streaks. The magnitude distributions for the supplementary catalogue
entries in the six passbands are shown in
Fig.~\ref{fig:supp_mags}. The dashed lines correspond to the nominal
brightness limits for read-out streak photometry and aperture
photometry for full-frame images early in the mission, so we would
expect the vast majority of sources to lie within the 1.5 mag interval
between the two lines. A small number of sources are found to be
brighter than the left dashed line because the bright limit has
increased slightly (by up to 0.2 mag depending on filter), over the
course of the {\em XMM-Newton} mission as the XMM-OM detector has
aged. Some spill-over of sources beyond the right dashed line is
expected because these are the faintest read-out streak measurements
and hence have the largest photmetric errors.  However, a tail of
sources is seen in several passbands between 0.5 and 2 magnitudes
fainter than the right dashed line. Stars with these magnitudes should
be well below the 0.97 counts/frame saturation limit of XMM-SUSS~2.1
by which they were selected for read-out-streak measurement. Therefore
we have visually inspected each of the sources which are more than 0.5
mag fainter than the dashed lines in Fig.~\ref{fig:supp_mags}, of
which there are 23 in total. We found that in 12 cases the bright
sources were on edge of the image, such that the count rate
measurement in XMM-SUSS~2.1 is flawed. The read-out-streak photometry
appears to be correct in 11 of the 12 cases. In a further 5 cases, the
sources are bright galaxies, with the XMM-SUSS~2.1 measurement
including counts from a much larger region than the standard
point-source aperture. In these cases the read-out-streak photometry
provides reasonable estimates for the nuclear regions of these
galaxies. The remaining 6 cases include 2 spurious scattered light
features which have been treated as extended sources in XMM-SUSS~2.1,
2 sources in which the presence of bright sources in the neighbouring
columns has caused problems with the local background estimate for the
read-out streak photometry, and 2 sources in which Flag number 11
appears to have been set incorrectly; in both these latter cases the
aperture photometry from the image agrees with the read-out streak
photometry. Overall the great majority of read-out streak measurements
appear to be correct among these apparently outlying points, so the
tail of faint read-out-streak magnitudes is not cause for concern.


\begin{table}
  \caption{Source statistics for the Supplementary Catalogue. 
The mean magnitudes for the sources in each filter are Vega magnitudes 
in the XMM-OM photometric system.}
\label{tab:supplementary}
\begin{tabular}{lccc}
Filter&Number of    &Number of    & Mean magnitude\\
      &saturated    &objects in   & in Supplementary\\
      &objects in   &Supplementary& Catalogue\\
      &XMM-SUSS~2.1 &Catalogue    & \\
\hline
V&978&258&11.06\\
B&1670&460&12.34\\
U&1252&440&11.35\\
UVW1&675&329&10.28\\
UVM2&88&53&8.80\\
UVW2&9&4&7.65\\
\hline
\end{tabular}
\end{table}

\subsection{Catalogue Access}
\label{sec:access}

The supplementary catalogue is available as a fits file via the MSSL
XMM-SUSS2 web
pages\footnote{http://www.mssl.ucl.ac.uk/mssl.astro/XMM-OM-SUSS/suss2$\_$1/}
and will be available from the ESA {\em XMM-Newton} Science Archive 
(XSA)\footnote{http://www.cosmos.esa.int/web/xmm-newton/xsa}.  

%
%
%

\section{Conclusions}
\label{sec:conclusions}

We have shown that read-out streaks in XMM-OM images can be used for photometric measurements of stars that are brighter than the coincidence-loss limit for normal aperture photometry. The study is based on XMM-OM $V$- and $B$-band measurements of stars in the Tycho-2 catalogue. We find that the recharge timescale for the pores of the microchannel plates in XMM-OM is 5.5$\pm0.1$ $\times 10^{-4}$~s, which sets the bright-source limit for read-out streak photometry to be 1.5 magnitudes brighter than the limit imposed by coincidence-loss in the CCD in full-frame images. We find that systematics and unaccounted errors limit the precision of the read-out streak photometry to 0.1 mag. As a demonstration, we have derived UVW1 photometry of the symbiotic nova RR Tel using read-out streaks, and compared it to photometry derived from earlier IUE observations. We have used the read-out streak method to construct a supplementary catalogue of photometry for sources which were too bright for photometric measurement in the XMM-SUSS2.1 catalogue. Using this method, photometry is recovered for 50 per cent of the UV measurements which exceeded the XMM-SUSS2.1 bright limit. 

\section*{Acknowledgments}
\label{sec:acknowledgments}

Based on observations obtained with XMM-Newton, an ESA science mission with instruments and contributions directly funded by ESA Member States and NASA.

\bibliographystyle{mn2e}

\label{lastpage}

\end{document}